\documentclass[twocolumn,showpacs,preprintnumbers,amsmath,amssymb]{revtex4}
\begin{document}
\draft

\title{On BPS preons, generalized holonomies and
$D=11$ supergravities}

\author{Igor A. Bandos$^{\dagger,\ast}$, Jos\'e A. de
Azc\'arraga$^{\dagger}$,
Jos\'e M. Izquierdo$^{\ddagger}$,
 Mois\'es Pic\'on$^{\dagger}$ and Oscar
Varela$^{\dagger}$}
\address{$^{\dagger}$Departamento de
F\'{\i}sica Te\'orica, Univ.~de Valencia and IFIC (CSIC-UVEG),
46100-Burjassot (Valencia), Spain
\\
$^{\ddagger}$Departamento de F\'{\i}sica Te\'orica,
 Facultad de Ciencias, Univ.~de Valladolid, 47011-Valladolid,
 Spain
\\
$^{\ast}$Institute for Theoretical Physics, NSC Kharkov Institute
of Physics  and Technology,  UA61108, Kharkov, Ukraine}

\def\theequation{\arabic{equation}}

\begin{abstract}
We develop the BPS preon conjecture to analyze the supersymmetric
solutions of $D=11$ supergravity. By relating the notions of
Killing spinors and BPS preons, we develop a moving $G$--frame
method ($G= GL(32,\mathbb{R})$, $SL(32,\mathbb{R})$ or
$Sp(32,\mathbb{R})$) to analyze their associated generalized
holonomies. As a first application we derive here the equations
determining the generalized holonomies of $\nu= k/32$
supersymmetric solutions and, in particular, those solving the
necessary conditions for the existence of BPS preonic ($\nu=
31/32$) solutions of the standard $D=11$ supergravity. We also
show that there exist elementary preonic solutions, {\it i.e.}
solutions preserving $31$ out of $32$ supersymmetries in a
Chern--Simons type supergravity. We present as well a family of
worldvolume actions describing the motion of pointlike and
extended BPS preons in the background of a D'Auria--Fr\'e type
$OSp(1|32)$--related supergravity model. We discuss the possible
implications for M-theory.

\end{abstract}

\pacs{11.30.Pb, 11.25.-w, 04.65.+e, 11.10.Kk; FTUV-03-1222
IFIC-03-60}

\maketitle

\narrowtext

\section{Introduction}

A complete, algebraic classification of M-theory BPS states, based
on the number $k$ of supersymmetries preserved by a given BPS
state, has been given in \cite{BPS01}. BPS states preserving $k$
out of $32$ supersymmetries are denoted $\nu=k/32$ states; $k=32$
corresponds to fully supersymmetric vacua. The observation
\cite{BPS01} (see also Sec.~\ref{secII}) that  BPS states that
break $32-k$ supersymmetries can be treated as composites of those
preserving all but one supersymmetries, suggests that the $k=31$
states might be considered as fundamental constituents of
M-theory. These $\nu=31/32$ BPS states were accordingly named {\it
BPS preons} \cite{BPS01}.

Interestingly enough, the notion of BPS preons may be extended to
arbitrary dimensions, including $D=4$. In this case, a point--like
BPS preon may be identified with a tower of massless higher spin
fields of all possible helicities (see \cite{BLS99,V01s},
\cite{30/32} and refs.~therein).

The actual existence of preonic, $\nu=31/32$ supersymmetric, BPS
states as solitonic solutions of the standard
Cremmer--Julia--Scherk (CJS) $D=11$ supergravity \cite{CJS} has
been the subject of recent studies \cite{Duff03,Hull03}. Although
no obstructions for their existence have been found by geometric
considerations based on the notion of generalized holonomy
\cite{Duff03} (see also \cite{FFP02,GP02,P+T03,P+T031}), no
$\nu=31/32$ solutions were found either.

In this paper we develop the notion of BPS preons to analyze the
various supersymmetric solutions of the supergravity equations. A
$\nu=k/32$ supersymmetric solution corresponds to a BPS state
composed of $n=(32-k)$ BPS preons. The corresponding $(32-k)$
bosonic spinors $\lambda_\alpha{}^r$ ($r=1, \ldots , n$) are
orthogonal to the $k$ Killing spinors $\epsilon_J{}^\alpha$
characterizing the $\nu=k/32$ solution,
\begin{eqnarray}
\epsilon_J{}^\alpha \, \lambda_\alpha{}^r & = & 0 \; , \\
\nonumber \alpha= 1, \ldots , 32\; , \quad J&=& 1, \ldots ,k\; ,
\quad r=1, \ldots ,(32-k) \; .
\end{eqnarray}
Thus, BPS preonic spinors and Killing spinors provide an
alternative (dual) characterization of a $\nu$-supersymmetric
solution; either one can be used and, for solutions with
supernumerary supersymmetries \cite{ppW,BDLW}, the
characterization provided by BPS preons is a more economic one.
Moreover, the use of both BPS preonic ($\lambda_\alpha{}^r$) and
Killing ($\epsilon_J{}^\alpha$) spinors allows us to develop (in
Sec.~\ref{secII}) a {\it moving G--frame} method, which may be
useful in the search for new supersymmetric solutions of CJS
supergravity.

We apply this moving $G$--frame method (in Sec.~\ref{secIII},
$G=SL(32,\mathbb{R})$) to studying the generalized holonomies of
CJS supergravity and discuss the basic equations characterizing
the still hypothetical BPS preonic solutions (Sec.~\ref{secIVB},
$G=GL(32,\mathbb{R})$, $SL(32,\mathbb{R})$ or
$Sp(32,\mathbb{R})$). Although no definite answer to the question
of the existence of BPS preonic solutions for the standard CJS
supergravity is given here, we do show (in Sec.~\ref{SubSecBPSCS})
that $\nu=31/32$ supersymmetric preonic configurations solve the
equations in the Chern-Simons (CS) supergravity case \cite{CS}
(for a review, see \cite{Zanelli}) {\it i.e.}, that CS
supergravity does have preonic solutions.

Using the recent results \cite{BdAI,BdAIL} on a gauge-fixed form
of the action for dynamical supergravity interacting with
dynamical superbrane sources (see Sec.~\ref{secIVC}), we also
propose (in Sec.~\ref{secIVD}) a $D=11$ worldvolume action for BPS
preons in the background of a D'Auria--Fr\'e $OSp(1|32)$--related
supergravity \cite{D'A+F} (see also \cite{CFGPN}), a model
allowing for an economic `embedding' of the standard $D=11$ CJS
supergravity.

In this paper we use a `mostly minus' metric,
$\eta_{ab}=(+,-,\ldots,-)$; the exterior derivative $d$ acts from
the right, $d\Omega_q = \frac{1}{q!} dx^{\mu_q} \wedge \cdots
\wedge dx^{\mu_1} \wedge dx^{\nu} \partial_\nu \Omega_{\mu_1
\cdots \mu_q}$.

\subsection{Equations of $D=11$ supergravity}\label{eqd11s}

The purely bosonic limit of the `free'
CJS supergravity equations is given by
\begin{eqnarray}\label{EiCJS}
E_{ab}&:=&  R_{ab} - {1\over 3} F_{a[3]}F_{b}{}^{[3]}
 + {1 \over 36} \eta_{ab}F_{[4]}F^{[4]}
=0\; ,
\\ \label{GFCJS}
{\cal G}_8 &:=&  d*F_4 - F_4\wedge F_4 = 0 \; ,
\\ \label{BIGFCJS}
dF_4 &\equiv & 0 \qquad
 \Leftrightarrow \qquad F_4=dA_3 \; ,
\end{eqnarray}
which include the Einstein equations with a contribution from the
energy--momentum tensor  of the antisymmetric gauge field
$A_{\mu\nu\rho}(x)$ only, Eq.~(\ref{EiCJS}), as well as the free
equations (\ref{GFCJS}) and Bianchi identities (\ref{BIGFCJS}) for
the gauge field in curved $D=11$ spacetime. Here $a, b, c
=0,\ldots , 9, 10\,= \mu, \nu , \rho$,
\begin{eqnarray}\label{A3}
A_3 &=& {1\over 3!} dx^\rho \wedge dx^\nu \wedge dx^\mu
A_{\mu\nu\rho}(x)\, , \nonumber \\ F_4 &=&  {1\over 4!} e^d\wedge
e^c\wedge e^b\wedge e^a F_{abcd}(x)  \; ,
\end{eqnarray}
$e^a=dx^\mu e_\mu{}^a(x)$,
$\, F_{a[3]}F^{b[3]}\equiv F_{ac_1c_2c_3}F^{bc_1c_2c_3}$, {\it etc.}
It is also assumed that the torsion and the gravitino vanish,
\begin{eqnarray}\label{T=0}
T^a &:=& De^a =de^a-e^b\wedge \omega_L{}_b{}^a =0\; , \quad
\\
\label{psi=0} \psi^\alpha &=& dx^\mu \psi_\mu^\alpha =0 \; ,
\qquad \alpha=1,\ldots , 32\; ,
\end{eqnarray}
where $\omega_L{}_b{}^a$ is the Lorentz connection. Such equations
possess {\it nonsingular} pp--wave solutions with supernumerary
supersymmetries \cite{ppW,BDLW}.

As the supergravity multiplet is the only one without higher spin
fields in $D=11$, no usual field-theoretical (spacetime) matter
contribution to the {\it r.h.s.}'s of Eqs. (\ref{EiCJS}),
(\ref{GFCJS}), (\ref{BIGFCJS}) may appear.  However, these
equations might be modified by higher order corrections in
curvature \cite{W00,Howe03} (a counterpart of the string
$\alpha^\prime$ corrections \cite{Zw85} in $D=10$, see also
\cite{HW96}), or/and by the presence of sources from $p$--branes.

The $\nu=16/32$ supersymmetric M2--brane solution of $D=11$
supergravity (see \cite{Duff94,Stelle88} and refs. therein)
possesses a singularity on the $(2+1)$-dimensional worldvolume
surface $W^{2+1}\subset M^{11}$, $x^\mu= \hat{x}^\mu (\xi) \equiv
\hat{x}^\mu (\tau, \sigma^1, \sigma^2)$ (a caret indicates a
function of the worldvolume coordinates). In other words, it
solves the Einstein field equation ${E}_{ab}={\cal T}_{ab} -
{1\over 9} \eta_{ab} \, {\cal T}_{c}{}^c$ with a singular
energy--momentum tensor density ${\cal T}_{ab} \propto \delta^3(x
-\hat{x}(\xi))$ \footnote{Notice that the tensor ${E}_{ab}$ in
(\ref{EiCJS}) is related to the {\it l.h.s.} of the true Einstein
equation of CJS supergravity, $\tilde{E}_{ab}$, by ${E}_{ab}=
\tilde{E}_{ab} - \frac{1}{D-2} \eta_{ab} \, \tilde{E}_{c}{}^c$.}.
The gauge field equation also possesses a singular contribution
$J_8$ in the {\it r.h.s.}, ${\cal G}_8 = J_8$, similar to that of
the electric current to the {\it r.h.s.}~of Maxwell equations. In
this sense, the M2--brane carries a supergravity counterpart of
the electric charge in Maxwell electrodynamics (see \cite{Duff91}
for a discussion).

The other basic $\nu=16/32$ supersymmetric solution of the CJS
supergravity, the M5-brane (see \cite{Gu92,Duff94,Stelle88} and
refs. therein), is a counterpart of Dirac monopole, {\it i.e.} of
the magnetically charged particle. It is characterized by a
modification of the Bianchi identities (Eq.~(\ref{BIGFCJS})) with
the analogue of a magnetic current in the {\it r.h.s.}, $dF_4 =
{\cal J}_5$.

The coincident M--brane solutions still possess $\nu=16/32$
supersymmetry, while the intersecting brane solutions correspond
to $\nu < 16/32$. Thus, as supernumerary supersymmetric solutions
($\nu > 1/2$) are known only for `free' CJS, it is reasonable to
consider first the `free' bosonic CJS equations
(\ref{EiCJS})--(\ref{BIGFCJS}), (\ref{T=0}), (\ref{psi=0}) in the
search for a hypothetical BPS preonic, $\nu=31/32$ solution.

Notice, however, that one should not  exclude the possible
existence of a brane solution [{\it i.e.} solutions of
${E}_{ab}={\cal T}_{ab} - {1\over 9} \eta_{ab} \, {\cal
T}_{c}{}^c$ with a ${\cal T}_{ab} \propto \delta^{p+1}(x
-\hat{x}(\xi))$] with supernumerary supersymmetries, although
certainly these solutions would describe quite unusual branes. The
reason why the `standard' brane solutions (like M--waves, M2 and
M5--branes in $D=11$) always break  $1/2$ of the supersymmetry is
that their $\kappa$--symmetry projector (the bosonic part of which
is identical to the projector defining the preserved
supersymmetries \cite{BKO,ST97}) has the form $(1-\bar{\Gamma})$
with $tr \bar{\Gamma}=0$, $\bar{\Gamma}^2=I$. However, worldvolume
actions for branes with a different form for the
$\kappa$--symmetry projector are known \cite{BL98,UZ,30/32}
although in an enlarged superspace (see \cite{JdA00}). A question
arises, whether such actions may be written in usual spacetime or
superspace.

As a partial answer to this question, we present here a $D=11$
spacetime action for a BPS preon in the background of a
D'Auria--Fr\'e type supergravity \cite{D'A+F}. The experience
provided by the usual $D=11$ M--branes and $D=10$ D-branes
together with the analysis \cite{BdAI,BdAIL} of the partial
preservation of local supersymmetry by the purely bosonic limit of
the super--$p$--brane action suggests that the existence of such a
preonic action implies that $\nu= 31/32$ solitons should exist in
a D'Auria--Fr\'e type model.

\subsection{Killing spinors, generalized connection
and generalized holonomy}

A bosonic  solution of the CJS supergravity equations preserving
$k$ out of $32$ supersymmetries can be characterized by $k$
independent bosonic spinors (Killing spinors),
$\epsilon_J{}^\alpha(x)$, $J=1,\ldots, k$, obeying the Killing
spinor equation
\begin{eqnarray}\label{Killing}
{\cal D} \epsilon_J{}^\alpha =d  \epsilon_J{}^\alpha -
\epsilon_J{}^\beta \omega_\beta{}^\alpha =0 \; . \qquad
\end{eqnarray}
The {\it generalized connection} $\omega_\beta{}^\alpha$ in
(\ref{Killing}) includes, besides the Lorentz (spin) connection
$\omega_L{}_\beta{}^\alpha= 1/4
\omega_L^{ab}\Gamma_{ab}{}_\beta{}^\alpha$, a tensorial part
$t_\beta{}^\alpha= \omega_\beta{}^\alpha-
\omega_L{}_\beta{}^\alpha$ constructed from the field strength
$F_{abcd}$, Eqs.~(\ref{BIGFCJS}), (\ref{A3}),
\begin{eqnarray}\label{CJSom}
 \omega_\beta{}^\alpha &=& {1\over 4}
\omega_L^{ab}\Gamma_{ab}{}_\beta{}^\alpha + {i\over 18} e^a
F_{ab_1b_2b_3} \Gamma^{b_1b_2b_3}{}_\beta{}^\alpha + \nonumber
\\
& + & {i\over 144} e^a \Gamma_{ab_1b_2b_3b_4}{}_\beta{}^\alpha
F^{b_1b_2b_3b_4} \; .
\end{eqnarray}

The Killing spinor equation (\ref{Killing}) comes from the
requirement of invariance under supersymmetry of the purely
bosonic solution (Eq.~(\ref{psi=0})) that requires
$\delta_{\varepsilon}\psi_\mu^\alpha = {\cal D}_\mu
\varepsilon^\alpha=0$. In $OSp(1|32)$--related  supergravity
models, including CS--type supergravities \cite{CS}, $ {\cal
D}_\mu$ and, hence, the Killing spinor equation involves a
$sp(32)$--valued $\omega_\beta{}^\alpha$ connection
($\omega^{\beta\alpha}:=C^{\beta\gamma}\omega_\gamma{}^\alpha=
\omega^{(\beta\alpha )}$) which is a true connection {\it i.e.},
it is associated with the actual gauge symmetry of the model.

In CJS supergravity, as well as in type IIB $D=10$ supergravity,
the gauge symmetry is restricted to $SO(1,10)$, and $\omega$ in
Eq.~(\ref{CJSom}) is not a true  connection. However, the
selfconsistency (integrability) condition for (\ref{Killing}),
${\cal D}{\cal D} \epsilon_J{}^\alpha =0$ has the suggestive form
\cite{FFP02,Duff03},
\begin{eqnarray}\label{eIR=0}
\epsilon_J{}^\beta  {\cal R}_\beta{}^{\alpha}  = 0\; ,
\end{eqnarray}
in terms of the {\it generalized curvature}
\begin{eqnarray}\label{calR}
{\cal R}_\beta{}^{\alpha} = d\omega_\beta{}^{\alpha} -
\omega _\beta{}^{\gamma}\wedge  \omega_\gamma{}^{\alpha}
\; .
\end{eqnarray}
Eqs.~(\ref{Killing}), (\ref{eIR=0}) {\it formally} possess a
$GL(32,\mathbb{R})$ gauge invariance with $\omega$ transforming as
a $GL(32,\mathbb{R})$ connection. However, $GL(32,\mathbb{R})$ is
not a gauge invariance of CJS supergravity, hence the name
`generalized' connection and curvature for $\omega$ and ${\cal
R}$.

Notice that, in contrast, a rigid $GL(k,\mathbb{R})$
transformation acting on the index $I$ results in a redefinition
of the Killing spinors $\epsilon_{I}{}^\alpha$ {\it i.e.}, in
replacing the Killing spinors by independent linear combinations
with constant coefficients. This is clearly allowed and
$GL(k,\mathbb{R})$ can be treated as a rigid symmetry of the
system of $k$ Killing spinors characterizing the $\nu=k/32$
supersymmetric solution of any model.

A true connection takes values in the Lie algebra ${\cal G}$ of
the {\it structure group} $G$ of the principal fiber bundle. The
curvature may take values in a smaller subalgebra ${\cal H}\subset
{\cal G}$ which is associated with a proper subgroup $H\subset G$
of the structure group, the {\it holonomy group}. For generalized
connections $\omega$ one may accordingly introduce the notion of
{\it generalized holonomy group} $H \subset G$ \cite{Duff03} (see
also \cite{FFP02,Hull03,P+T03}), such that ${\cal
R}_\beta{}^{\alpha} \in {\cal H}$ (while $\omega_\beta{}^{\alpha}
\in {\cal G}$). In this light, the necessary condition for the
existence of $k$ Killing spinors, Eq. (\ref{eIR=0}), can be
treated as a restriction on  the generalized holonomy group $H$
\cite{Duff03,Hull03,P+T03,P+T031,BDLW}.

It has been shown that for both CJS $D=11$ supergravity
\cite{Hull03} and for type IIB supergravity \cite{P+T031}, the
generalized holonomy group $H$ is a subgroup of
$SL(32,\mathbb{R})$, $H\subset SL(32,\mathbb{R})$. As the
generalized connections are clearly traceless in both cases, one
also has $G\subset SL(32,\mathbb{R})$.

For $OSp(1|32)$--related models including CS supergravities
\cite{CS} the (true) structure group is $G\subset
Sp(32,\mathbb{R})$ and the holonomy group is $H\subset
Sp(32,\mathbb{R})$.

A full expression for the generalized curvature ${\cal
R}_{\alpha}{}^{\beta}$ corresponding to purely bosonic solutions
of CJS supergravity may be found {\it e.g.}, in \cite{BEWN83,GP02}
(see Appendix B of \cite{GP02} and refs. therein; there,
$\eta_{ab}$ and $F$ correspond to $-\eta_{ab}$, $-2F$). For our
purposes here it is sufficient to note that this ${\cal
R}_{\alpha}{}^{\beta}$ obeys
\begin{widetext}
\begin{eqnarray}\label{GiRCJS=}
i_a {\cal R}_{\alpha}{}^{\gamma}\Gamma^a{}_{\gamma}{}^{\beta} &=&
- {1\over 4} e^b R_{b[c_1 c_2c_3]} \Gamma^{c_1c_2c_3} + {1\over 2}
e^a \, E_{ab} \Gamma^b_{\alpha}{}^{\beta} + \nonumber \\
&+&  {i\over 36}  e^a\, (\Gamma_{a}{}^{b_1b_2b_3}
+ 6 \delta_{a}^{[b_1} \Gamma^{b_2b_3]})_{\alpha}{}^{\beta}
\, [*{\cal G}_8]_{b_1b_2b_3}
+ {i\over 720} \, e^a\, [dF_4]_{b_1\ldots b_5}\,
(\Gamma_{a}{}^{b_1\ldots b_5} +
10  \delta_{a}^{[b_1} \Gamma^{b_2\ldots b_5]})_{\alpha}{}^{\beta}
\; . \qquad
\end{eqnarray}
\end{widetext}
where $E_{ab}$, ${\cal G}_8$ are the {\it r.h.s}'s of the Einstein
and the gauge field equations as defined in Eqs. (\ref{EiCJS}),
(\ref{GFCJS}) and $i_a$ is defined by   $i_a e^b=\delta_a^b$ so
that {\it i.e.}, for $\Omega_p={1\over p!} e^{a_p} \wedge \ldots
\wedge e^{a_1}\Omega_{a_1\ldots a_p}$,
\begin{equation}
i_a\Omega_p={1\over (p-1)!} e^{a_{p}} \wedge \ldots \wedge
e^{a_2}\Omega_{aa_2\ldots a_{p}} \quad ;
\end{equation}
in particular $i_a {\cal R}_{\alpha}{}^{\beta}=e^b {\cal R}_{ab
\alpha}{}^\beta$. The equality (\ref{GiRCJS=}) implies that the
set of the {\it free bosonic} (Eq.~(\ref{psi=0})) equations for
the CJS supergravity, Eqs. (\ref{EiCJS}), (\ref{GFCJS}),
(\ref{BIGFCJS}), is equivalent to the following simple equation
for the generalized curvature of Eq.~(\ref{calR}), $e^b {\cal
R}_{ab\alpha}{}^\gamma \Gamma^a{}_\gamma{}^\beta=0$ or
\begin{equation}\label{EqCJS=0}
i_a {\cal R}_{\alpha}{}^{\gamma}\; \Gamma^a{}_{\gamma}{}^{\beta}
=0\; ,
\end{equation}
since the {\it r.h.s}~of Eq.~(\ref{GiRCJS=}) is zero on account of
the equations of motion (\ref{EiCJS}), (\ref{GFCJS}), the Bianchi
identity (\ref{BIGFCJS}) and that $R_{b[c_1 c_2c_3]} =0$ [since
$R_{abcd}=R_{cdab}$ and  $DT^a=0$ by Eq.~(\ref{T=0})].

\section{Killing spinors, preons and generalized
$G$--frame}\label{secII}

\subsection{Killing spinors and BPS states}

A BPS state $\vert BPS\, , \, k\rangle$  described by a solitonic
solution preserving $k$ supersymmetries is, schematically,  one
satisfying
\begin{eqnarray}\label{kSUSY}
\epsilon_J{}^\alpha  Q_\alpha \vert BPS\, , \, k\rangle = 0 \, ,
\quad J=1,\ldots, k\; ,  \quad k \leq 31 \; ,
\end{eqnarray}
where $Q_\alpha$ are the supersymmetry generators obeying
\begin{eqnarray}\label{QQP}
 \{ Q_\alpha , Q_\beta\} &=& P_{\alpha \beta} \; ,
\quad [Q_\alpha , P_{\beta\gamma}]=0 \\
\alpha , \beta , \gamma &=& 1,2, \ldots, 32 \quad , \nonumber
\end{eqnarray}
so that $P_{\alpha \beta}= P_{\beta\alpha}$. The generalized
momentum $P_{\alpha \beta}$ may be decomposed {\it e.g.}  in the basis of
$D=11$ $Spin(1,10)$ ($32 \times 32$)
 Dirac matrices,
\begin{eqnarray}\label{P32}
P_{\alpha\beta} = P_\mu \Gamma^\mu_{\alpha\beta} + Z_{\mu\nu}
\Gamma^{\mu\nu}_{\alpha\beta} +  Z_{\mu_1\ldots \mu_5}
\Gamma^{\mu_1\ldots \mu_5}_{\alpha\beta}\; ,
\end{eqnarray}
giving then rise to the standard $D=11$ momentum $P_\mu$ and to
the tensorial `central' charge  generators $Z_{\mu\nu}$,
$Z_{\mu_1\ldots \mu_5}$ of the M--algebra $\{Q_\alpha , Q_\beta \}
= P_{\alpha \beta}$. These generators may be identified as
topological charges \cite{JdAT} related to the M2-- and M5--branes
(as well as to the M9--brane and KK7--brane of M-theory
\cite{Hull97}; see \cite{ST97} for the r\^ole of the worldvolume
gauge fields of the M5--brane, and \cite{HH,JdA00} for D-branes).

\subsection{BPS preons as constituents}

A BPS preon \cite{BPS01} state $\vert BPS\, ,\, 31 \rangle \equiv
\vert \lambda \rangle$
is a state characterized by a single bosonic spinor
$\lambda_\alpha$,
\begin{eqnarray}\label{preon}
P_{\alpha \beta} \vert \lambda \rangle = \lambda_\alpha
\lambda_\beta \vert \lambda \rangle \; .
\end{eqnarray}
(hence the notation $\vert \lambda \rangle$) or as a state
preserving  all supersymmetries but one (hence the notation $\vert
BPS , 31 \rangle$) \cite{BPS01}. The bosonic spinor parameters
$\epsilon_I{}^\alpha$ corresponding to the supersymmetries
preserved by a BPS preon $|\lambda \rangle$,
\begin{eqnarray}\label{preonSUSY}
& \epsilon_I{}^\alpha  Q_\alpha \vert \lambda \rangle = 0 \;  ,
\qquad I=1,\ldots,  31 \; ,
\end{eqnarray}
are `orthogonal' to the bosonic spinor $\lambda_\alpha$ that
labels it,
\begin{eqnarray}\label{eIl=0}
& \epsilon_I{}^\alpha \lambda_\alpha =0 \; , \qquad I=1,\ldots, 31
\;
\end{eqnarray}
(see below). Identifying $\epsilon_I{}^\alpha$ with $31$ Killing
spinors satisfying (\ref{Killing}), one finds that (\ref{eIl=0})
expresses the fact that these Killing spinors are orthogonal to
the {\it single bosonic spinor $\lambda_\alpha$ characterizing a
hypothetical  BPS preonic solution}.

For BPS states $\vert BPS, k \rangle$ preserving $k\leq 30$
supersymmetries (\ref{kSUSY})
 one may introduce $n=32-k$ bosonic spinors
$\lambda_\alpha{}^r$ corresponding to the $n$ broken
supersymmetries. They may be treated as characterizing $n$ BPS
preons $\vert \lambda{}^r\rangle$, $r=1,\ldots , n$, out of which
the corresponding $k/32$ BPS state is composed \cite{BPS01}.

\medskip

 To make this transparent, let us consider the eigenvalue
matrix $p_{\alpha\beta}$ of the generalized momentum operator
$P_{\alpha\beta}$  corresponding to the BPS state $\vert BPS, k
\rangle$ (which is usually assumed to be an eigenstate of the
generalized momentum, {\it i.e.} having definite energy and
definite brane charges), $P_{\alpha\beta} \vert BPS, k \rangle =
p^{(k)}_{\alpha\beta} \vert BPS, k \rangle$. This is a symmetric
matrix of $rank (p_{\alpha\beta}^{(k)})= n=32-k$ (a relation
justified below). Hence $p_{\alpha\beta}^{(k)}$ may be
diagonalized by a general linear transformation {\it i.e.},
\begin{equation}\label{p=GpG}
p_{\alpha\beta}^{(k)}= g_{\alpha}{}^{(\gamma)}p_{(\gamma)(\delta)}
g_{\beta}{}^{(\delta)}
\end{equation}
 with $p_{(\gamma)(\delta)}= diag (\ldots )$.
Moreover, as $ g_{\alpha}{}^{(\gamma)}\in GL(n,\mathbb{R})$, this
diagonal matrix can be put in the form
\begin{equation}\label{pdiag0}
p_{(\gamma)(\delta)}= diag (\underbrace{1,\ldots,1,-1,\ldots, -1}_{n=32-k},
\underbrace{0,\ldots,0}_{k})\; ,
\end{equation}
where the number of nonvanishing elements, all $+1$ or $-1$, is
equal to $n=rank (p_{\alpha\beta}^{(k)})$. However, the usual
assumptions for the supersymmetric quantum mechanics describing
BPS states do not allow for negative eigenvalues of
$P_{\alpha\beta}=\{Q_\alpha , Q_\beta\}$ [$p_{11}=-1$, e.g., would
imply $(Q_1)^2 \vert BPS, k \rangle = - \vert BPS, k \rangle$,
contradicting positivity]. Thus, only positive eigenvalues are
allowed and
\begin{equation}\label{pdiag}
p_{(\gamma)(\delta)}= diag (\underbrace{1,\ldots,1}_{n=32-k},
\underbrace{0,\ldots,0}_{k})\; .
\end{equation}

Substituting (\ref{pdiag}) into (\ref{p=GpG}), one
arrives at
\begin{equation}\label{p=G1G}
p_{\alpha\beta}^{(k)}= g_{\alpha}{}^{(\gamma)}\;
\left(\begin{matrix}
   \begin{matrix}
    1           & &
    \cr
          &
 \ddots
                  &
     \cr
                & &  1
         \end{matrix}
                          &  \hbox{\LARGE $0$} \cr
\hbox{\LARGE $0$} &
   \begin{matrix}
    0           & &
    \cr
          &
               \ddots
                   &
     \cr
                & &  0
         \end{matrix}
\end{matrix}
\right)_{(\gamma)(\delta) } \;\; g_{\beta}{}^{(\delta )} \; ,
\end{equation}
or, equivalently, denoting $ g_{\alpha}{}^{1}=
 \lambda_{\alpha}{}^{1}$, $\ldots$, $ g_{\alpha}{}^{n}=
 \lambda_{\alpha}{}^{n}$,
{\setlength\arraycolsep{0pt}\begin{eqnarray}\label{npreon}
P_{\alpha\beta}&\vert BPS, k \rangle& = \sum\limits_{r=1}^{n=32-k}
\lambda_{\alpha}{}^r \lambda_{\beta}{}^r \vert BPS, k \rangle \; \equiv \nonumber \\
&\equiv& \left(\lambda_{\alpha}{}^1\lambda_{\beta}{}^1 + \ldots +
\lambda_{\alpha}{}^n \lambda_{\beta}{}^n \right) \vert BPS, k
\rangle \; . \qquad
\end{eqnarray}}\\[-8pt]
One sees using Eq.~(\ref{QQP}) that, if the preserved
supersymmetries correspond to the generators $\epsilon_J{}^\alpha
Q_\alpha$, $J=1, \ldots , k$, Eq. (\ref{kSUSY}), then
\begin{equation} \sum\limits_{r=1}^{n=32-k} \epsilon_{(J}{}^\alpha
\lambda_{\alpha}{}^r\; \epsilon_{K)}{}^\beta \lambda_{\beta}{}^r =
0\; ,
\end{equation}
which,  immediately implies
\begin{eqnarray}\label{eKl=0}
 & \epsilon_{J}{}^\alpha \lambda_{\alpha}{}^r=0 \; , \\
 & J = 1,...,k\; , \quad r=1,..., n \quad , \nonumber
\end{eqnarray}
making clear that $k=32-n$. This explains the relation $n=32-k$
between the number of preons $n= rank(p_{\alpha\beta})$ and the
number of preserved supersymmetries $k$. The $k=31$ case ($n=1$)
is Eq. (\ref{eIl=0}) for BPS preons.

Eq. (\ref{npreon}) may be looked at as a manifestation of the {\it
composite structure} of the $\nu = k/32$ BPS state $\vert BPS, k
\rangle$,
\begin{eqnarray}\label{k=npreon}
\vert BPS, k \rangle = \vert \lambda^1 \rangle \otimes \ldots
\otimes  \vert \lambda^{(32 - k)} \rangle \; ,
\end{eqnarray}
where $ \vert \lambda^1 \rangle$, $\ldots$, $ \vert
\lambda^n\rangle$ are BPS elementary, preonic states characterized
by the spinors $\lambda_\alpha{}^1$ , $\ldots$,
$\lambda_\alpha{}^n$, respectively.

\subsection{Moving $G$-frame}

When a BPS state $\vert k \rangle$ is realized as a solitonic
solution of supergravity, it is characterized by $k$ Killing
spinors $\epsilon_J{}^\beta (x)$ or by the $n=32-k$ bosonic
spinors $\lambda_\alpha{}^r (x)$ associated with the $n$ BPS
preonic components of the state $\vert BPS, k \rangle$. The
Killing spinors and the preonic spinors are orthogonal,
\begin{eqnarray}\label{eIlr=0}
 \epsilon_J{}^\alpha \lambda_\alpha{}^r &=& 0 \, ,
\nonumber \\   J=1,\ldots,  k \, ,&& \; r =1,\ldots,  n=32-k \; .
\end{eqnarray}
and, hence, may be completed to obtain bases in the spaces of
spinors with upper and with lower indices by introducing $n=32-k$
spinors $w_r{}^\alpha$  and $k$ spinors $u_{\alpha}{}^L$
satisfying
\begin{eqnarray}\label{wl=}
 w_s{}^\alpha \lambda_\alpha{}^r= \delta_s^r \; , \quad
 w_s{}^\alpha  u_\alpha{}^J =0 \; , \quad
 \epsilon_J{}^\alpha  u_\alpha{}^K =  \delta_J{}^K \; .
\end{eqnarray}
Either of these two dual bases defines a {\it generalized moving
$G$--frame} described by the nondegenerate matrices
\begin{eqnarray}\label{g}
g_{\alpha}{}^{(\beta)} = \left( \lambda_\alpha{}^s \, ,
u_\alpha{}^J \right)\; , \qquad
g^{-1}{}_{(\beta)}{}^\alpha = \left( \begin{array}{c} w_s{}^\alpha \\
\epsilon_J{}^\alpha \end{array} \right)\; , \quad
\end{eqnarray}
where $(\alpha)=(s,J)=(1,\ldots,32-k;J=1,\ldots,k)$. Indeed,
$g^{-1}{}_{(\beta)}{}^\gamma g_{\gamma}{}^{(\alpha)} =
\delta_{(\beta)}{}^{(\alpha)}$ is equivalent to Eqs. (\ref{wl=})
and (\ref{eIlr=0}), while
\begin{eqnarray}\label{g-1g}
\delta_{\alpha}{}^{\beta} = g_{\alpha}{}^{(\gamma)}
g^{-1}{}_{(\gamma)}{}^\beta \equiv  \lambda_\alpha{}^r w_r{}^\beta
+ u_\alpha{}^J \epsilon_J{}^\beta \;
\end{eqnarray}
provides the unity decomposition or
completeness relation in terms of these dual bases.

One may consider the dual basis $g^{-1}{}_{(\beta)}{}^\alpha$ to
be constructed from the bosonic spinors in
$g_{\alpha}{}^{(\beta)}$ by solving Eq.~(\ref{g-1g}) or
$g^{-1}g=I$ (Eqs. (\ref{wl=}) and (\ref{eIlr=0})). Alternatively,
one may think of $w_r{}^\alpha$ and $u_\alpha{}^J$ as being
constructed from $\epsilon_J{}^\alpha$ and $\lambda_\alpha{}^r$
through a solution of the same constraints. In this sense {\it the
generalized moving $G$--frame (\ref{g}) is constructed from $k$
Killing spinors $\epsilon_J{}^\alpha$ characterizing the
supersymmetries preserved by a BPS state (realized as a solution
of the supergravity equations) and from the $n=32-k$ bosonic
spinors $\lambda_\alpha{}^r$ characterizing the BPS preons from
which the BPS state is composed}.

Although many of the considerations below are general, we shall be
mainly interested here in the cases $G=SL(32,\mathbb{R})$ and
$G=Sp(32,\mathbb{R})$.

Clearly, in $D=11$, the charge conjugation matrix
$C^{\alpha\beta}=- C^{\beta\alpha}$ allows us to express the dual
basis  $g^{-1}$ in terms of the original one $g$ or {\it vice
versa}. In particular, in the preonic $k=31$ case one finds that,
as $\lambda_\alpha C^{\alpha\beta}\lambda_\beta \equiv 0$,
$\lambda^\alpha= C^{\alpha\beta}\lambda_\beta$ has to be expressed
as $\lambda^\alpha= \lambda^I \epsilon_I{}^\alpha$, for some
coefficients $\lambda^I$, $I=1,\ldots, 31$. In general (as {\it
e.g.}, in CJS supergravity with nonvanishing $F_4$), the charge
conjugation matrix is not `covariantly constant', ${\cal
D}C^{\alpha\beta}= -2 \omega^{[\alpha\beta]} \not= 0$. This
relates the coefficients $\lambda^I= \lambda^\alpha u_\alpha{}^I$
to the antisymmetric (non--symplectic) part of the generalized
connection,
$\omega^{[\alpha\beta]}=C^{[\alpha\gamma}\omega_\gamma{}^{\beta]}$
by $d\lambda^I - A \lambda^I= 2 \lambda_\alpha
\omega^{[\alpha\beta]}u_\beta{}^I$ \footnote{To see this, one
calculates $d\lambda^I={\cal D}\lambda^I= ({\cal
D}C^{\alpha\beta})\lambda_\beta u_\alpha{}^I+C^{\alpha\beta}({\cal
D}\lambda_\beta)u_\alpha{}^I +C^{\alpha\beta}\lambda_\beta {\cal
D}u_\alpha{}^I$ and use Eq.~(\ref{Dl31}), (\ref{Du31}) to find
$d\lambda^I=A\lambda^I+2\lambda_\alpha
\omega^{[\alpha\beta]}u_\beta{}^I$}. In $OSp(1|32)$--related
models $\omega^{[\alpha\beta]}=0$ and $A=0$, hence $\lambda^I$ is
constant and we may set $\lambda^I=\delta^I_{31}$ using the {\it
global} transformations of $GL(31,\mathbb{R})$, which is a rigid
symmetry of the system of Killing spinors . This allows us to
identify $\lambda^\alpha$ itself with one of the Killing spinors
\begin{eqnarray}\label{OSpe}
G= Sp(32,\mathbb{R})\; : \;  \epsilon_I{}^\alpha &=&
(\epsilon_i{}^\alpha, \lambda^\alpha)\; , \quad \lambda^\alpha:=
C^{\alpha\beta}\lambda_\beta\; , \qquad \\ \nonumber
 &&  i=1,\ldots , 30 \; .
\end{eqnarray}

Without specifying a solution of the constraints (\ref{g-1g}) (or
$g^{-1}g=I$), the moving frame possesses a $G=GL(32,\mathbb{R})$
symmetry. One may impose as  additional constraints $\mathrm{det}
(g)=1$ or $\mathrm{det}(g^{-1})=1$ reducing $G$ to
$SL(32,\mathbb{R})$,
\begin{eqnarray}\label{SLG}
G= SL(32,\mathbb{R})\; :  \quad \mathrm{det}(g_\beta^{\,
(\alpha)})=1= \mathrm{det} (g^{-1}_{\; (\alpha)}{}^\beta)\; .
\end{eqnarray}
 For instance, in the
preonic case $k=31$ this would imply $w^\alpha =1/(31)!
\varepsilon^{\alpha\beta_1\ldots \beta_{31}} u_{\beta_1}{}^1
\ldots u_{\beta_{31}}{}^{31}$.  Such a frame is most convenient to
study the   bosonic solutions of CJS supergravity.

\section{Geometry of BPS preons and $\nu = k/32$--supersymmetric
solutions}\label{secIII}

\subsection{Generalized connection and moving $G$--frame}

The Killing equation (\ref{Killing}) for a $\nu=k/32$
supersymmetric solution,
\begin{eqnarray}\label{Killingk}
{\cal D} \epsilon_J{}^\alpha =d  \epsilon_J{}^\alpha -
\epsilon_J{}^\beta \omega_\beta{}^\alpha =0 \; , \qquad
J=1,\ldots, k\; ,
\end{eqnarray}
implies the following equations for the other components of the
moving $G$--frame
\begin{eqnarray}\label{Dlk}
{\cal D} \lambda_{\alpha}{}^r &:=& d\lambda_{\alpha}{}^r +
\omega_{\alpha}{}^{\beta} \, \lambda_{\beta}{}^r =
\lambda_{\alpha}{}^s \, A_s{}^r \; , \qquad
\\ \label{Duk}
{\cal D} u_\alpha{}^J  &:=&
 du_\alpha{}^J + \omega_{\alpha}{}^\beta
u_\beta{}^J = \lambda_{\alpha}{}^r \, B_r^J \; , \qquad
\\ \label{Dwk}
{\cal D} w_r{}^\alpha  &:=& dw_r{}^\alpha - w_r{}^\beta
\omega_{\beta}{}^{\alpha} = - A_r{}^s w_s{}^\alpha - B_r^J
\epsilon_J{}^\alpha \; ,   \qquad \\ \nonumber &&  \qquad \alpha ,
\beta = 1, \ldots, 32; \\ \nonumber \quad J &=& 1,\ldots, k \; ;
\qquad r,s=1,\ldots, (32-k) \; ,
\end{eqnarray}
where $A_s{}^r$ and $B_r{}^I$ are $(32-k) \times (32-k)$ and
$(32-k) \times k$ arbitrary one--form matrices.

To obtain Eqs.~(\ref{Dlk}), (\ref{Duk}), (\ref{Dwk}) one can take
firstly the  derivative ${\cal D}$ of  the orthogonality relations
(\ref{eIlr=0}), (\ref{wl=}).  After using Eq.~(\ref{Killingk}),
this results in
\begin{eqnarray}\label{Dort1}
\epsilon_I{}^\alpha {\cal D}\lambda_\alpha{}^r=0 &,&
\epsilon_I{}^\alpha {\cal D}u_\alpha{}^J=0\; , \qquad
\\ \label{Dort2}
w_s{}^\alpha {\cal D}\lambda_\alpha{}^r = - {\cal D} w_s{}^\alpha
\; \lambda_\alpha{}^r  &,&  w_s{}^\alpha {\cal D}u_\alpha{}^J = -
{\cal D} w_s{}^\alpha \, u_\alpha{}^J\, . \qquad
\end{eqnarray}
Then, for instance, to derive  (\ref{Dlk}), one uses the unity
decomposition (\ref{g-1g}) to express ${\cal D}\lambda_\alpha{}^r$
through the contractions $w_s{}^\alpha {\cal D}\lambda_\alpha{}^r$
and $\epsilon_I{}^\alpha {\cal D}\lambda_\alpha{}^r$: $\;  {\cal
D}\lambda_\alpha{}^r \equiv \lambda_\alpha{}^s\, w_s{}^\beta {\cal
D}\lambda_\beta{}^r + u_\alpha{}^I \, \epsilon_I{}^\beta {\cal
D}\lambda_\beta{}^r$. The second term vanishes due to
(\ref{Dort1}), while the first one is not restricted by the
consequences of the Killing spinor equations and may be written as
in Eq.~(\ref{Dlk}) in terms of an arbitrary form $A_s{}^r \equiv
w_s{}^\alpha {\cal D}\lambda_\alpha{}^r$.

Notice that, using the unity decomposition (\ref{g-1g}),  one may
also solve formally  Eqs. (\ref{Killingk}), (\ref{Dlk}),
(\ref{Duk}), (\ref{Dwk}) with respect to the generalized
connection $\omega_\alpha{}^\beta$,
\begin{eqnarray}\label{om=A+B}
\omega_\alpha{}^\beta &=& A_r{}^s \, \lambda_{\alpha}{}^r
w_s{}^\beta + B_r{}^J \lambda_{\alpha}{}^r  \epsilon_J{}^\beta
-(dg g^{-1})_\alpha{}^\beta \; ,
\end{eqnarray}
where $g_{\alpha}^{\, (\beta)}$ and $g^{-1}_{\;
(\beta)}{}^{\alpha}$ are defined in Eq. (\ref{g}) and, hence,
\begin{equation}\label{dgg-1G}
(dg g^{-1})_\alpha{}^\beta = d\lambda_{\alpha}{}^r \, w_r{}^\beta
+ du_\alpha{}^I \, \epsilon_I{}^\beta  \; . \qquad
\end{equation}

For a BPS $\nu= 31/32$, preonic configuration Eqs. (\ref{Dlk}),
(\ref{Duk}), (\ref{Dwk}) read
\begin{eqnarray}\label{Dl31}
{\cal D} \lambda_{\alpha} &:=& d\lambda_{\alpha} +
\omega_{\alpha}{}^\beta \lambda_{\beta} = A \lambda_{\alpha} \; ,
\qquad
\\ \label{Du31}
{\cal D} u_\alpha{}^I &:=& du_\alpha{}^I + \omega_{\alpha}{}^\beta
u_\beta{}^I = B^I \lambda_{\alpha} \; , \qquad
\\ \label{Dw31}
{\cal D} w^\alpha  &:=& dw^\alpha - w^\beta
\omega_{\beta}{}^{\alpha} = - A\,  w^\alpha - B^I
\epsilon_I{}^\alpha \;   \qquad
\end{eqnarray}
and contain $1+31=32$ arbitrary one--forms  $A$ and $B^I$.

For $G=SL(32,\mathbb{R})$ one may choose $det(g)=1$,
Eq.~(\ref{SLG}), which implies $tr(dg g^{-1}):= (dg
g^{-1})_\alpha{}^\alpha =0$. Then the $sl(32,\mathbb{R})$--valued
generalized connection $\omega_\alpha{}^\beta$
($\omega_\alpha{}^\alpha =0$) allowing for a $\nu=k/32$
supersymmetric configuration is determined by Eq.~(\ref{om=A+B})
with $A_r{}^r=0$,
\begin{eqnarray}\label{trA=0}
G= SL(32,\mathbb{R})\; :  \qquad A_r{}^r=0\; .
\end{eqnarray}

In particular, the $sl(32,\mathbb{R})$--valued generalized
connection allowing for a BPS preonic, $\nu=31/32$, configuration,
should have the form
\begin{eqnarray}\label{om31=B}
G= SL(32,\mathbb{R}) \; , &&  \nu=31/32 \; : \quad
\nonumber \\
\omega_\alpha{}^\beta &=&  B^I\; \lambda_{\alpha}
\epsilon_I{}^\beta - (dg g^{-1})_\alpha{}^\beta \;  \qquad
\end{eqnarray}
in terms of   $31$ arbitrary one-forms $B^I$, $I=1, \ldots , 31$.

Assuming a definite form of the generalized connections, {\it
e.g.} the one characterizing bosonic solutions of the `free' CJS
supergravity equations (\ref{CJSom}), one finds that
Eqs.~(\ref{om=A+B}) become differential equations for $k$ Killing
spinors $\epsilon_{J}{}^\alpha$ {\it and $n=32-k$ BPS preonic
spinors $\lambda_\alpha{}^r$} once $(dg
g^{-1})=d\lambda_{\alpha}{}^r w_r{}^\beta - u_\alpha{}^I
d\epsilon_I{}^\beta$ (Eq.~(\ref{dgg-1G})) is taken into account.

On the other hand, one might reverse the argument and ask for the
structure of a theory allowing for $\nu=k/32$ supersymmetric
solutions. This question is especially interesting for the case of
BPS preonic and $\nu=30/32$ solutions as, for the moment, such
solutions are unknown in the standard $D=11$ CJS and $D=10$ type
II supergravities.

\subsection{Generalized holonomy for BPS preons and for $\nu =
k/32$ supersymmetric solutions}

The simplest application of the moving $G$--frame construction
above is to find an explicit form for the general solution of
Eq.~(\ref{eIR=0}), which expresses the necessary conditions for
the existence of $k$ Killing spinors. As the Killing spinor
equation (\ref{Killingk}) implies Eqs. (\ref{Dlk}), (\ref{Duk}),
one may solve instead the selfconsistency conditions for these
equations,
\begin{eqnarray}\label{DDlk}
{\cal D}{\cal D}\lambda_\alpha{}^r &=& {\cal R}_\alpha{}^{\beta}
\lambda_\beta{}^r =  \lambda_\alpha{}^s (dA-A\wedge A)_s{}^r \\
 \label{DDuk}
{\cal D}{\cal D}u_\alpha{}^I &=& {\cal R}_\alpha{}^{\beta}
u_\beta{}^I =  \lambda_\alpha{}^r (dB_r^I + B_s^I \wedge
A_r{}^s)\; .
\end{eqnarray}
 Using the unity decomposition (\ref{g-1g}), which implies
${\cal R}_\alpha{}^{\beta}= {\cal R}_\alpha{}^{\gamma}
\lambda_\gamma{}^r \, w_r{}^\beta + {\cal R}_\alpha{}^{\gamma}
u_\gamma{}^I \, \epsilon_I{}^\beta$,
 one finds the following expression for the generalized curvature
\begin{equation}\label{kcalR=gl}
 {\cal R}_\alpha{}^\beta = G_r{}^s \, \lambda_{\alpha}{}^r w_s{}^\beta
+ \nabla B_r^I \lambda_{\alpha}{}^r  \epsilon_I{}^\beta \; ,
\end{equation}
where
\begin{eqnarray} \label{Grs=}
 && G_r{}^s :=  (dA - A\wedge A)_r{}^s \; , \qquad \\
\label{nbB=}
&& \nabla B_r^I:= dB_r^I -
A_r{}^s \wedge B_s^I \; ,
\end{eqnarray}
For $k=31$, corresponding to the case of a BPS preon,
Eq.~(\ref{kcalR=gl}) simplifies to
\begin{equation}\label{calR=gl}
 {\cal R}_\alpha{}^\beta = dA \, \lambda_{\alpha} w^\beta
+(dB^I + B^I \wedge A) \,  \lambda_{\alpha}  \epsilon_I{}^\beta \;
.
\end{equation}
Eqs.~(\ref{kcalR=gl}) and (\ref{calR=gl}) imply ${\cal
R}_\alpha{}^\beta = \lambda_\alpha{}^r (\cdots)_r{}^\beta$ and,
thus, clearly solve  Eq.~(\ref{eIR=0}), $ \epsilon_I{}^\beta {\cal
R}_\beta{}^{\alpha}  = 0$.

The conditions $G\subset SL(32,\mathbb{R})$ and hence $H \subset
SL(32,\mathbb{R})$, ${\cal R}_\alpha{}^\alpha=0$ (which is always
the case for bosonic solutions of `free' CJS and type IIB
supergravities \cite{Hull03,P+T031}), imply  $A_r{}^r=0$ in
Eq.~(\ref{kcalR=gl}) [see Eq.~(\ref{trA=0})], while for $k=31$
Eq.~(\ref{calR=gl}) simplifies to
\begin{eqnarray}\label{calR=sl}
 H\subset SL(32,\mathbb{R})\; , &\;& k=31 \;  : \;  \nonumber \\
& & {\cal R}_\alpha{}^\beta = dB^I  \lambda_{\alpha}
\epsilon_I{}^\beta \; . \qquad
\end{eqnarray}

Finally, for $G\subset Sp(32,\mathbb{R})$ $\omega^{[\alpha\beta
]}=0$, the holonomy group $H\subset Sp(32,\mathbb{R})$,  $ {\cal
R}^{\alpha\beta} := C^{\alpha\gamma}{\cal R}_\gamma{}^\beta =
{\cal R}^{(\alpha\beta )}$, and Eq.~(\ref{calR=sl}) reduces to
 \begin{eqnarray}\label{calR=sp}
 H\subset Sp(32,\mathbb{R})\; , &\;&  k=31 \;  : \; \nonumber \\
& & {\cal R}_\alpha{}^\beta = dB  \, \lambda_{\alpha}
\lambda^\beta \; , \qquad
\end{eqnarray}
where only one arbitrary one--form $B$ appears [to obtain
(\ref{calR=sp}) one has to keep in mind that $\epsilon_I{}^\alpha
= (\epsilon_i{}^\alpha, C^{\alpha\beta}\lambda_\beta)$,
$I=(i,31)$, Eq.~(\ref{OSpe})]. {\it Eqs.~(\ref{calR=sl}),
(\ref{calR=sp}) solve Eq.~(\ref{eIR=0}) for preons when
$G=SL(32,\mathbb{R}), \ Sp(32,\mathbb{R})$, respectively}.

Eq.~(\ref{kcalR=gl}) with $A_r{}^r=0$ (Eq.~(\ref{trA=0}), and,
hence,  $(dA - A\wedge A)_r{}^r=0$) provides an explicit
expression for the results of \cite{Hull03,P+T031} on generalized
holonomies of $k$--supersymmetric solutions of $D=11$ and of
$D=10$ type IIB supergravity, namely $\; H\subset \;
SL(32-k,\mathbb{R})$ $\subset\!\!\!\!\!\!\times
\mathbb{R}^{k(32-k)}$. For a BPS preon $k=31$, and $\; H\subset
{\mathbb{R}}^{31}$ as expressed by Eq. (\ref{calR=sl}). However,
our explicit expressions for the
($sl(32-k,\mathbb{R})\,\subset\!\!\!\!\!\! +
\mathbb{R}^{k(32-k)}$)--valued generalized curvatures ${\cal
R}_\alpha{}^\beta$, Eqs.~(\ref{kcalR=gl}), (\ref{calR=sl}), given
in terms of the Killing spinors $ \epsilon_I{}^\beta$ and bosonic
spinors $\lambda_{\alpha}{}^r$ characterizing the BPS preon
contents of a $\nu=k/32$ BPS state, may be useful in searching for
new supersymmetric solutions, including preonic $\nu=31/32$ ones.
Some steps in this direction are taken in the next section.

\section{On BPS preons in $D=11$ CJS supergravity and
beyond}

\subsection{BPS preons in Chern--Simons supergravity}
\label{SubSecBPSCS}

The first
observation is that the generalized curvature allowing for a
BPS preonic
($k=31$ supersymmetric) configuration for the case of
 $H\subset SL(32,\mathbb{R})$ holonomy, Eq. (\ref{calR=sl}),
is nilpotent
\begin{eqnarray}\label{RR=0}
{\cal R}_\alpha{}^\gamma \wedge {\cal R}_\gamma{}^\beta =0
\;,\quad \mathrm{for} \quad H\subset SL(32,\mathbb{R}) \;\; , \;
k=31 \;. \qquad
 .
\end{eqnarray}
As a result it solves the purely bosonic equations of a
Chern--Simons supergravity [see \cite{CS,Zanelli}, although our
statement may be related to a more general version of a
hypothetical Chern--Simons--like supergravity],
\begin{eqnarray}\label{RRRRR=0}
{\cal R}_\alpha{}^{\gamma_1} \wedge {\cal
R}_{\gamma_1}{}^{\gamma_2} \wedge {\cal R}_{\gamma_2}{}^{\gamma_3}
\wedge {\cal R}_{\gamma_3}{}^{\gamma_4} \wedge {\cal
R}_{\gamma_4}{}^\beta =0\; .
\end{eqnarray}

Clearly, the same is true for $H\subset Sp(32,\mathbb{R}) \subset
SL(32,\mathbb{R})$, where ${\cal R}$ is given by Eq.
(\ref{calR=sp}). Thus, there exist BPS preonic solutions in CS
supergravity theories, including $OSp(1|32)$-type ones.

Note that Eq. (\ref{RR=0}) follows in general for a preonic
configuration only. For the configurations preserving $k\leq 30$
of the $32$ supersymmetries, the bosonic equations of a CS
supergravity, Eqs. (\ref{RRRRR=0}) reduce to (see (\ref{Grs=}),
(\ref{nbB=}))
\begin{eqnarray}\label{GGGGG=0}
G_{s}{}^{s_2} \wedge G_{s_2}{}^{s_3} \wedge G_{s_3}{}^{s_4} \wedge
G_{s_4}{}^{s_5} \wedge G_{s_5}{}^{r} =0 \;\; , \nonumber \\
G_{s}{}^{s_2} \wedge G_{s_2}{}^{s_3} \wedge G_{s_3}{}^{s_4} \wedge
G_{s_4}{}^{r} \wedge \nabla B_{r}{}^{I} =0 \;\; ,
\end{eqnarray}
which are not satisfied identically for $G_r{}^r=0$.
Eqs.~(\ref{GGGGG=0}) are satisfied {\it e.g.}, by configurations
with $G_s{}^r=0$, for which the generalized holonomy group is
reduced down to $H \subset \mathbb{R}^{\otimes k(32-k)}$, ${\cal
R}_\beta{}^\alpha= \nabla B^I_r \lambda_\beta{}^r
\epsilon_{I}{}^{\alpha}$.

Thus, {\it only} the preonic, $\nu=31/32$, configurations {\it
always} solve the Chern--Simons supergravity equations
(\ref{RRRRR=0}).

\subsection{Searching for preonic solutions of `free' bosonic
CJS equations}\label{secIVB}

We go back now to the question of whether BPS $\nu=31/32$
(preonic) solutions exist for the standard CJS supergravity
\cite{CJS}. As it was noted in Sec.~\ref{eqd11s}, this problem can
be addressed step by step, beginning by studying the existence of
preonic solutions for `free' bosonic CJS equations. To this aim it
is useful to observe \cite{GP02} that these equations  may be
collected in a compact expression for the generalized curvature,
Eq. (\ref{EqCJS=0}). The generalized curvature of a BPS preonic
configuration satisfies Eq. (\ref{calR=sl}), and thus it solves
the `free' bosonic CJS supergravity equations (\ref{EqCJS=0}) if
 \begin{eqnarray}\label{GldB=0}
i_adB^I \, \epsilon_{_I}{}^\alpha \Gamma^a{}_{\alpha}{}^{\beta}
=0\; .
\end{eqnarray}
Actually, Eq.~(\ref{calR=sl}) in Eq.~(\ref{EqCJS=0}) gives
$\lambda_\alpha \, i_a dB^I \, \epsilon_I{}^\gamma
 \Gamma^a{}_{\gamma}{}^{\beta}  =0$. However, as
 $\lambda_\alpha\not=0$, this is equivalent to Eq.~(\ref{GldB=0}).

Eq. (\ref{GldB=0}) contains a summed $I=1, \ldots , 31$ index and,
as a result, it is not easy to handle. It would be much easier to
deal with the expression $\Gamma^a{}_{\alpha}{}^{\gamma}i_{a}
{\cal R}_{\gamma}{}^{\beta}$ which, with Eq.~(\ref{calR=sl}) is
equal to $\Gamma^a{}_{\alpha}{}^{\gamma} \lambda_{\gamma}i_a dB^J
 \epsilon_{J}{}^{\beta}$. Indeed, $(\Gamma^a \lambda)_{\alpha}i_a dB^J
 \epsilon_{J}{}^{\beta}=0$, for instance, would imply
 $(\Gamma^a \lambda)_{\alpha}i_a dB^J =0$ which may be shown to have
 only trivial solutions. However,
$\Gamma^a{}_{\alpha}{}^{\gamma} i_a {\cal R}_\gamma{}^{\beta} \neq
 0$ in general {\it for a solution of the `free'
bosonic CJS equations} (Eq.~(\ref{EqCJS=0})),
\begin{equation}\label{GaiacalR=}
\Gamma^a{}_{\alpha}{}^{\gamma}i_a{\cal R}_{\gamma}{}^{\beta} = -
{i \over 12} \left( D\hat{F}_{\alpha}{}^{\beta}  + \, {\cal O}
(F\, F) \right) \; , \quad
\end{equation}
where $D=e^aD_a$ is the Lorentz covariant derivative [not to be
confused with ${\cal D}$ defined  in Eqs.~(\ref{Killing}),
(\ref{CJSom})],
\begin{eqnarray}\label{hatF=}
\hat{F}_{\alpha}{}^{\beta} = F_{a_1a_2a_3a_4}
(\Gamma^{a_1a_2a_3a_4})_{\alpha}{}^{\beta}
\; , \qquad
\end{eqnarray}
and $ {\cal O} (F\, F)$ denotes the terms of second order in
$F_{c_1c_2c_3c_4}$,
\begin{widetext}
\begin{eqnarray}\label{OFF}
 {\cal O} (F\, F)&=& {1\over (3!)^2 \, 4!}\, e^a (\Gamma_a{}^{b_1b_2b_3} +
2 \delta_a^{[b_1}\Gamma^{b_2b_3]})
\epsilon_{b_1b_2b_3[4][4^\prime]} F^{[4]}F^{[4^\prime]} +  \\
\nonumber &+&  {2i\over 3} e^a (\Gamma_a{}^{b_1b_2b_3b_4} + 3
\delta_a^{[b_1}\Gamma^{b_2b_3b_4]}) F_{cdb_1b_2}F^{cd}{}_{b_3b_4}
+ {8i\over 9} e^aF_{abb_1b_2}F^{b}{}_{b_3b_4b_5}
\Gamma^{b_1b_2b_3b_4b_5} \; .
\end{eqnarray}
\end{widetext}

Eq.~(\ref{calR=sl}) then implies that for a hypothetical preonic
solution of the `free' bosonic CJS equations, the gauge field
strength $F_{abcd}$ should be nonvanishing (otherwise $dB^J=0$ and
${\cal R}_\alpha{}^\beta=0$, see above) and satisfy
\begin{eqnarray}\label{GdB31=}
\Gamma^a{}_{\alpha}{}^{\gamma} \lambda_{\gamma}\; i_adB^J \;
\epsilon_{J}{}^{\beta}= - {i \over 12}
\left(D\hat{F}_{\alpha}{}^{\beta}
+ \, {\cal O} (F\, F)\right) \; . \qquad
\end{eqnarray}
Using (\ref{wl=}), Eqs. (\ref{GdB31=}) split into a set of
restrictions for $F_{abcd}$,
\begin{eqnarray}\label{DFl=0}
\left( D\hat{F} + \, {\cal O} (F\, F)\right){}_{\alpha}{}^{\beta}
\lambda_{\beta} =0 \; ,
\end{eqnarray}
and equations for $dB^I$,
\begin{eqnarray}\label{GdB31=u}
\Gamma^a{}_{\alpha}{}^{\gamma}
\lambda_{\gamma}\; i_adB^I \; = - {i \over 12} \left( D\hat{F}  +
\, {\cal O} (F\, F)
\right){}_{\alpha}{}^{\beta}  \; u_\beta{}^I .
\qquad
\end{eqnarray}
{\it Eq.~(\ref{GdB31=}) or, equivalently,
Eqs.~(\ref{DFl=0}),(\ref{GdB31=u}) are the equations to be
satisfied by a CJS preonic configuration}. Note that if a
nontrivial solution of the above equations with some
$F_{abcd}\not=0$ and some $dB^I\not=0$ is found, one would have
then to check in particular that such a solution satisfies $ddB^I=
0$ and $D_{[e}F_{abcd]}=0$.

On the other hand, if the general solution of the above equation
turned out to be trivial, $dB^I = 0$, this would imply ${\cal
R}_\alpha{}^\beta =0$ and, thus, a trivial generalized holonomy
group, $H=1$. However, this is the necessary condition for fully
supersymmetric, $k=32$, solutions \cite{FFP02}. Hence a trivial
solution for Eqs. (\ref{DFl=0}), (\ref{GdB31=u}) would indicate
that a solution preserving $31$ supersymmetries possesses all $32$
ones (thus corresponding to a fully supersymmetric vacuum) and,
hence, that there are no preonic, $\nu=31/32$ solutions of the
{\it free bosonic} CJS supergravity equations (\ref{EiCJS}),
(\ref{GFCJS}), (\ref{BIGFCJS}), (\ref{T=0})  and (\ref{psi=0}).

If this happened to be the case, one would have to study the
existence of preonic solutions for the CJS supergravity equations
with nontrivial right hand sides. These could be produced by
corrections of higher--order in curvature \cite{W00,Howe03,HW96}
(a counterpart of the string $\alpha^\prime$ corrections in $D=10$
\cite{Zw85}) and by the presence of sources (from some exotic
$p$--branes).

\subsection{On brane solutions and worldvolume
actions}\label{secIVC}

As far as supersymmetric $p$--brane solutions of supergravity
equations are concerned, one notices that for most of the known
$\nu=1/2$ supersymmetric solutions ($\nu =16/32$ in $D=11$ and
$D=10$ type II cases) there also exist worldvolume actions in the
corresponding ($D=11$ or $D=10$ type II) superspaces possessing
$16$ $\kappa$-symmetries, exactly the number of supersymmetries
preserved by the supergravity solitonic solutions. The {\it
$\kappa$--symmetry--preserved supersymmetry}  correspondence was
further discussed and extended for the case of $\nu < 1/2$
multi-brane solutions in \cite{BKO,ST97}.

In this perspective one may expect that if  preonic $\nu=31/32$
supersymmetric solutions of the CJS equations with a source do
exist,  a worldvolume action possessing $31$ $\kappa$--symmetries
should also exist in a curved $D=11$ superspace. For a moment no
such actions are known in the {\it standard} $D=11$ superspace,
but they are known in a superspace enlarged with additional
tensorial `central' charge coordinates \cite{BL98,UZ}. One might
expect that the r\^ole of these additional tensorial coordinates
could be taken over by the tensorial fields of $D=11$
supergravity. But this would imply that the corresponding action
does not exist in the flat standard $D=11$ superspace as it would
require a contribution from the above additional field degrees of
freedom (replacing the tensorial coordinate ones corresponding to
spirit of \cite{JdA00}). This lack of a clear flat {\it standard}
superspace limit hampers the way towards a hypothetical
worldvolume action for a BPS preon in the usual curved $D=11$
superspace.

Nevertheless, a shortcut in the search for such an action may be
provided by the recent observation \cite{BdAIL} that the
superfield description of the dynamical supergravity--superbrane
interacting system, described by the sum of the {\it superfield}
action for supergravity (still unknown for $D=10,11$) and the
super--$p$--brane action, is gauge equivalent to the much simpler
dynamical system described by the sum of the spacetime, component
action for supergravity and the action {\it for the purely bosonic
limit} of the super--$p$--brane. This bosonic $p$--brane action
carries the memory of being the bosonic limit of a
super--$p$--brane by still possessing $1/2$ of the spacetime local
supersymmetries \cite{BdAI}; this preservation of local
supersymmetry reflects the $\kappa$--symmetry of the original
super--$p$--brane action.

Thus the $\kappa$-symmetric worldvolume actions for
super--$p$--branes have a clear spacetime counterpart: the purely
bosonic actions in spacetime possessing a part of local spacetime
supersymmetry of a `free' supergravity theory.

This fact, although explicitly discussed for the standard,
$\nu=1/2$ superbranes in \cite{BdAIL}, is general since it follows
from symmetry considerations only and thus it applies to any
superbrane, including a hypothetical preonic one. The number of
supersymmetries possessed by this bosonic brane action coincides
with the number of $\kappa$--symmetries of the parent
super--$p$--brane action. Moreover, these supersymmetries are
extracted by a projector which may be identified with the bosonic
limit of the $\kappa$--symmetry projector for the superbrane. With
this guideline in mind one may simplify, in a first stage, the
search for a worldvolume action for a BPS preon in standard
supergravity (or in a model minimally extending the standard
supergravity) by discussing the bosonic limit that such a
hypothetical action should have.

\subsection{BPS preons in D'Auria--Fr\'e
supergravity}\label{secIVD}

Let us consider a symmetric spin--tensor one--form
$e^{\alpha\beta} = e^{\beta\alpha} = dx^\mu e_\mu^{\alpha\beta}
(x)$ transforming under local supersymmetry by
\begin{eqnarray}\label{eabsusy}
\delta_{\varepsilon} e^{\alpha\beta} = - 2 i \psi^{(\alpha} \,
\varepsilon^{\beta)} \; ,
\end{eqnarray}
where $\psi^{\alpha}$ is a fermionic one--form,
\begin{eqnarray}\label{psiF}
 \psi^{\alpha} =  dx^\mu \psi_\mu^{\alpha}(x)\; ,
\end{eqnarray}
which we may identify with the gravitino. Let us consider for
simplicity the worldline action ({\it cf. } \cite{BL98})
\begin{eqnarray}\label{BPSac}
S &=& \int\limits^{}_{W^1} \lambda_\alpha (\tau)\, \lambda_\beta
(\tau) \hat{e}^{\alpha\beta} \nonumber \\
&=&  \int d\tau \lambda_\alpha (\tau)\, \lambda_\beta (\tau) \,
e_\mu^{\alpha\beta} (\hat{x}(\tau)) \,
\partial_\tau \hat{x}^\mu (\tau)
\; ,
\end{eqnarray}
where $\tau$ parametrizes the worldline $W^1$ in $D=11$ spacetime,
$\hat{e}^{\alpha\beta}:= d\tau \partial_\tau \hat{x}^\mu (\tau) \,
e_\mu^{\alpha\beta} (\hat{x}(\tau))$ and $\lambda_\alpha (\tau)$
is an auxiliary spinor field on the worldline $W^1$. The extended
($p\ge 1$) object counterpart of this worldline action is the
following action for tensionless $p$--branes (cf.~\cite{UZ})
\begin{eqnarray}
S_{p+1}&=& \int\limits^{}_{W^{p+1}} \lambda_\alpha \lambda_\beta
\hat{\rho} \wedge
\hat{e}^{\alpha\beta} \nonumber\\
&=& \int\limits^{}_{W^{p+1}} d^{p+1} \xi \, \rho^k  \lambda_\alpha
\lambda_\beta \hat{e}^{\alpha\beta}_\mu \partial_k \hat{x}^\mu \;,
\label{Sp+1}
\end{eqnarray}
where $\hat{\rho}(\xi)$ is a $p$-form auxiliary field, and $\rho^k
(\xi)$ is the worldvolume vector density (see \cite{BZ}) related
to $\hat{\rho}(\xi)$ by $\hat{\rho}(\xi)= (1/p!) d\xi^{j_p} \wedge
\ldots \wedge d\xi^{j_1} \rho_{j_1\ldots j_p}(\xi) =(1/p!)
d\xi^{j_p} \wedge \ldots \wedge d\xi^{j_1} \epsilon_{j_1\ldots
j_pk} \rho^k(\xi)$.

One easily finds that the action (\ref{BPSac}) possesses all but
one of the local spacetime supersymmetries \footnote{Notice that
when a brane action is considered in a supergravity {\it
background}, the local spacetime supersymmetry is not a gauge
symmetry of that action but rather a transformation of the
background; it becomes a gauge symmetry only when a supergravity
action is added to the brane one so that supergravity is
dynamical.}, Eq.~(\ref{eabsusy}), $31$ for $\alpha, \beta=1,
\ldots, 32$ corresponding to $D=11$. Indeed, performing a
supersymmetric variation $\delta_\varepsilon$ of (\ref{BPSac})
assuming $\delta_{\varepsilon} \lambda_\alpha (\tau)=0$, one finds
\begin{eqnarray}\label{susyBPSac}
\delta_{\varepsilon} S &=& - 2i \int\limits^{}_{W^1}
\hat{\psi}^\alpha \lambda_\alpha (\tau)\;
\hat{\varepsilon}^\beta
\lambda_\beta (\tau)  \; .
\end{eqnarray}
Thus, one sees that $\delta_{\varepsilon} S=0$ for the
supersymmetry parameters on $W^1$ that obey ({\it cf. }
(\ref{eIl=0}))
\begin{eqnarray}\label{el=0W}
\hat{\varepsilon}^\beta \lambda_\beta (\tau) =0  \quad
(\hat{\varepsilon}^\beta:= {\varepsilon}^\beta (\hat{x}(\tau))) \;
.
\end{eqnarray}
Clearly Eq. (\ref{el=0W}) possesses $31$ solutions, which may be
expressed through worldvolume spinors
$\hat{\epsilon}_I{}^\alpha(\tau)$ (the worldline counterparts of
the Killing spinors) orthogonal to $\lambda_\alpha (\tau)$,
$\hat{\epsilon}_I{}^\alpha(\tau)\lambda_\alpha (\tau)=0$, as
\begin{eqnarray}\label{ve=eI}
\hat{\varepsilon}^\beta = {\varepsilon}^I(\tau)
\hat{\epsilon}_I{}^\beta \quad , \quad I=1,\ldots,31 \quad ,
\end{eqnarray}
for some arbitrary ${\varepsilon}^I(\tau)$. The same is true for
the tensionless $p$-branes described by the action (\ref{Sp+1}).

Thus, the actions (\ref{BPSac}), (\ref{Sp+1}) possess $31$ of the
$32$ local spacetime supersymmetries (\ref{eabsusy}) and, in the
light of the discussion  of the previous section, can be
considered as the spacetime counterparts of a superspace
BPS--preonic action (hypothetical in the standard superspace but
known \cite{BL98,BLS99,UZ} in flat maximally enlarged or tensorial
superspaces).

The question that remains to be settled is the meaning of the
symmetric spin--tensor one--form $e^{\alpha\beta}$ with the local
supersymmetry transformation rule (\ref{eabsusy}) in $D=11$
supergravity. The contraction of $e^{\alpha\beta}$ with
$\Gamma_a$,
\begin{eqnarray}\label{ea=}
e^a = e^{\alpha\beta} \Gamma^a_{\alpha\beta} \quad ,
\end{eqnarray}
may be identified with the $D=11$ vielbein. Decomposing
$e^{\alpha\beta}$ in the basis of the $D=11$ $Spin (1,10)$
gamma--matrices,
\begin{eqnarray}\label{eab=}
e^{\alpha\beta} &=& e^{\beta\alpha}= \qquad \\ \nonumber &=&
{1\over 32} e^a \Gamma_a^{\alpha\beta} - \frac{1}{2! \,32}
B_1^{ab} \Gamma_{ab}^{\alpha\beta} + \frac{1}{5! \, 32}
B_1^{a_1\ldots a_5} \Gamma_{a_1\ldots a_5}^{\alpha\beta}\; ,
\end{eqnarray}
one finds that $e^{\alpha\beta}$ also contains the antisymmetric
tensor one--forms $B_1^{ab}(x)=dx^\mu B_\mu^{ab}(x)$ and
$B_1^{a_1\ldots a_5}(x)= dx^\mu B_\mu^{a_1\ldots a_5}(x)$. Such
fields, moreover, with exactly the same supersymmetry
transformation rules, are involved in the $D=11$ supergravity
model by D'Auria and Fr\'e \cite{D'A+F}.

Thus the action (\ref{BPSac}) can be teated as a worldline action
for a BPS preon in the background of the D'Auria--Fr\'e,
$OSp(1|32)$--related `gauge' supergravity model. This might be
regarded as an indirect indication of the existence of BPS preonic
$\nu=31/32$ solutions in D'Auria--Fr\'e $D=11$ supergravity.

The possibility of having preonic actions in the standard CJS
supergravity requires additional study.

\section{Discussion and Outlook}

In this paper we have studied  the r\^ole of the BPS preon notion
\cite{BPS01} in the analysis of supersymmetric solutions of $D=11$
supergravity. This notion suggests the moving $G$--frame method,
which we propose as a new tool in the search for supersymmetric
solutions of $D=11$ and $D=10$ supergravity. We used this method
here to make a step towards answering whether the standard CJS
supergravity \cite{CJS} possesses a solution preserving $31$
supersymmetries, a solution that would correspond to a BPS preon
state. Although this question has not been settled for the CJS
supergravity case, we have shown in our framework that preonic,
$\nu=31/32$ solutions do exist in a Chern--Simons type $D=11$
supergravity \cite{CS}.

Although the main search for preonic solutions concerns the `free'
bosonic CJS supergravity equations, one should not exclude other
possibilities, both inside and outside the CJS standard
supergravity framework. When, {\it e.g.}, super--$p$--brane
sources are included, the Einstein equation (\ref{EiCJS}), and
possibly the gauge field equations (\ref{GFCJS}) and even the
Bianchi identities (\ref{BIGFCJS}), acquire {\it r.h.s}'s. In this
case (see Eq.~(\ref{GiRCJS=})), a {\it r.h.s.} also appears in Eq.
(\ref{EqCJS=0}) and the situation would have to be reconsidered.
Another source of modification of the CJS supergravity equations
might be due to `radiative' corrections of higher order in
curvature. Such modified equations might also allow for preonic
solutions not present in the unmodified ones. If it were found
that only the inclusion of these higher--order curvature terms
allows for preonic BPS solutions, this would indicate that BPS
preons cannot be seen in a classical low energy approximation of
M-theory and, hence, that they are intrinsically quantum objects.

The special r\^ole of BPS preons in the algebraic classification
of all the M-theory BPS states \cite{BPS01} allows us to
conjecture that they are elementary (quark-like) necessary
ingredients of any model providing a more complete description of
M-theory. In such a framework, if the standard supergravity did
not contain $\nu =31/32$ solutions, neither in its `free' form,
nor in the presence of a super--$p$--brane source, this might just
indicate the need for a wider framework for an effective
description of M--theory. Such an approach could include
Chern--Simons supergravities \cite{CS} and/or the use of larger,
extended superspaces (see \cite{JdA00,JdAIMO} and refs.~therein),
in particular with additional tensorial coordinates (also relevant
in the description of massless higher--spin theories
\cite{BLS99,V01s}). In this perspective our observation that the
BPS preonic configurations do solve the bosonic equations of
Chern--Simons supergravity models looks interesting.

{\it Note added}. We mention that it might be interesting to look
at the role of vectors and higher order tensors that may be
constructed from the preonic spinors $\lambda_\alpha{}^r$, in
analogy with the use of the Killing vectors $K_{IJ}^a=\epsilon_I
\Gamma^a \epsilon_J$ and higher order bilinears $\epsilon_I
\Gamma^{a_1\cdots a_s} \epsilon_J$ made in Refs.
\cite{GP02,GoNeWa03,H-JPaSm03,GaGuPa03}.

\medskip

{\it Acknowledgments}. This work has been partially supported by
the research grants BFM2002-03681, BFM2002-02000 from the
Ministerio de Ciencia y Tecnolog\'{\i}a and from EU FEDER funds,
by the grant N 383 of the Ukrainian State Fund for Fundamental
Research, the INTAS Research Project N 2000-254, and by the Junta
de Castilla y Le\'on grant A085-02. M.P. and O.V. wish to thank
the Ministerio de Educaci\'on, Cultura y Deporte and the
Generalitat Valenciana, respectively, for their FPU and FPI
research grants.

\end{document}